\documentclass[letterpaper]{JHEP}

\usepackage{epsfig}
\usepackage{graphicx}

\preprint{SUGP-00/6-2\\
 hep-th/0007171}

\title{Half-Branes, Singular Brane Intersections, and Kaluza-Klein reduction}
\author{Donald Marolf\\Physics Department, Syracuse University, Syracuse,
         New York 13244}
\date{May, 2000}
\abstract{Two subtle aspects of brane intersections are investigated.
The first concerns the `half-branes' that arise in discussions of
the Hanany-Witten effect, often in the D0/D8 setting.
The second involves the validity of
seemingly singular classical BPS brane intersections.
A study of
holomorphic curves in the background of
a Kaluza-Klein monopole and the associated reduction to type IIA
supergravity sheds light on both issues.
Many seemingly singular D2/D6 intersections are shown
to lift to smooth configurations of M2-branes in 11-dimensions, and
a mechanism is found for certain
$Z_2$ confinement effects in type II string theories that eliminates
any need for half-branes.
}
\begin{document}

\section{Introduction}
In recent years the
study of intersecting branes, especially of the BPS sort, has provided fertile
ground
for the development of ideas in
string theory, gauge theory, and gravity.  For example,
BPS intersecting brane
scenarios provided the setting for 
the accounting of black hole entropy in the case studied by Strominger
and Vafa \cite{SV} and for the subsequent generalizations.
In addition, the Hanany-Witten framework \cite{HW} and the
Maldacena conjecture \cite{juan} have forged strong links between gauge theories
and brane intersections.  As a result,
the construction and study of intersecting brane solutions has become a minor
industry.

The case of intersections with D6-branes is particularly tractable due to the
fact that a unit charged D6-brane in ten dimensions lifts to the Kaluza-Klein
monopole of 11-dimensional supergravity.  This solution is just the usual
4+1 Kaluza-Klein monopole \cite{KK} times a flat
6-dimensional space.  The important
property is that the unit charged monopole solution is completely smooth, so
that it can be approximated by flat Minkowski space at its center.
More generally, the `core' of the multi-charged Kaluza-Klein monopole is
just an orbifold singularity.
This means that the `near-core' versions of certain
ten-dimensional supergravity
solutions involving intersections with D6-branes can be constructed
\cite{ITY,Aki,GM}
by an appropriate quotient of branes in 11-dimensional Minkowski
space.
We note that for appropriately chosen parameters (large internal $S^1$), the
curvature of
a unit charged Kaluza-Klein monopole is everywhere small, so that a description
in terms
of classical supergravity is appropriate.

When a full supergravity solution is not available, one can often extract
useful
information by considering one brane as a `test' or `probe' brane in a
background
spacetime determined by another brane (see, e.g., \cite{Ima,CGS,NOYY,Yosh,BG}
and others in the intersecting
brane context).  This description is appropriate when the
test brane is much lighter than the background brane.  
When derivatives on the brane
are appropriately small (see, \cite{DBI} for a review),
static configurations of the test branes
are determined by the Dirac-Born-Infeld action. 
In the BPS case, the Dirac-Born-Infeld action may lead to
BPS solutions even in the presence of large curvatures \cite{Thor}.  

In this context,
D6-branes again provide a particularly tractable setting.  For example, a test
D2-brane
in the background of a D6-brane is described in M-theory as a test M2-brane in
the
background of a Kaluza-Klein monopole.   The point here is that the
Kaluza-Klein
manifold is K\"ahler (in fact, hyper-K\"ahler).  In a static
spacetime of the form $K \times R$
where $K$ is K\"ahler and $R$ is the time direction, static BPS configurations
of M2-branes
are exactly described by holomorphic curves in $K$.  This simplifying property
was
used in \cite{NOYY,Yosh,BG} to study the Hanany-Witten brane-creation effect.

Here, we continue the study of test D2-branes intersecting
D6-branes by mapping out
in detail the associated charge distributions in ten-dimensions.
Our goal here is two-fold.  First, we wish to investigate the
issue of half-branes
that arises naturally in discussions of the Hanany-Witten effect, e.g.
\cite{HW,PS,Lif,BDG,DFK,BGL,K,HoWu,dA,WB,BGS,LeaI,LeaII}. 
Consider for example a system with a single D0-brane and a single D8-brane and
suppose that the boundary conditions are such that the ten-form Ramond-Ramond
gauge field takes the symmetric values $\pm 1/2$ of the fundamental quantum
on either side of the D8-brane
domain wall.  Then, certain arguments \cite{PS,BGL} (and 
\cite{Lif,BDG,DFK,K,HoWu,dA,OSZ} in related contexts)  in massive
type IIA supergravity \cite{Romans} lead to the conclusion
that exactly 1/2 of a fundamental string must end on the D0-brane.  While this
seems to be at odds with charge quantization, several possible resolutions
immediately present themselves.  One possibility is that the half-string is
a mere artifact of some accounting scheme 
(see, e.g. \cite{BDS,Taylor,Mor,3Q,SS}) 
and that it
is not in fact in conflict with charge quantization.  Another
possibility is that such symmetric boundary conditions for 
D8-branes are not in fact allowed,
and that the Ramond-Ramond gauge ten-form field strength $F_{10}$
must in fact take integer values.
A final possibility is that $F_{10}$ is in fact allowed to take half-integer
values but that, in such backgrounds, D0-brane charge is allowed to occur
only in multiples of 2.  By considering the T-dual D2/D6 system, we will
uncover evidence in support of this final alternative of $Z_2$ confinement
of D0-brane charge.

Our second goal is to investigate in detail the case where
the branes actually meet and intersect.  The point here is that the
background metric and/or the dilaton 
is typically singular at the location of the background
brane.  Thus, a priori, the Dirac-Born-Infeld effective action is not an
adequate description of
the test brane at the point where it
actually intersects the background brane\footnote{In the particular case
of the Neveu-Schwarz 5-brane there is an exact conformal field theory in
this region with which one can work, see e.g. \cite{EGKRS, Pelc}.}.  
However, as mentioned above, the
corresponding
core region is smooth for the unit charge Kaluza-Klein monopole.  Thus,
studying smooth
test M2-branes in 11-dimensions should tell us about the allowed intersections
of D2- and D6-branes in ten dimensions.  As we will see, some of these
intersections
look quite singular from the ten-dimensional point of view.  In particular, we
find
that many solutions analogous to the fractional string solutions of \cite{CGS}
correspond to smooth M2-branes in 11-dimensions.  Nevertheless, the proper
string
charge always appears in integral quanta.

The plan of our paper is as follows.  In section \ref{jred}, we consider
a holomorphic curve in the Kaluza-Klein monopole background
which has been previously discussed \cite{BG} in connection
with the Hanany-Witten effect and compute its reduction to the type IIA
D6-background.  Although the corresponding D2-brane does not in fact intersect
the
D6-brane, it illustrates a number of effects and provides some
insight
into the question of half-branes.
The associated charge distributions are described in
terms of the three notions of charge reviewed in \cite{3Q}: brane source
charge, Maxwell charge, and Page charge \cite{Page}.  This will show that
certain
fractional brane issues in the D2/D6-context are resolved
(as in the D0/D2-brane case of \cite{BDS,Taylor,Mor,SS}, see also
\cite{3Q}) by proper accounting techniques.  However, we will also
see that, in this particular example, the system cannot be compactified along
the D2-brane.  As
a result,
T-duality makes no connections with the case of the D0/D8-system.

Section \ref{hol} then considers holomorphic curves for which the corresponding
D2-branes do intersect the D6-brane.
In fact, we consider the most general smooth holomorphic curves whose reduction
to
type IIA yields rotationally symmetric D2-branes.  One particular class
of such D2-branes has an analogue in the compactified system.   
Curves in this class
feature an essentially flat D2-brane connected to the D6-brane by a thin tube
that approximates a collection of fundamental strings in much the same way
as the BIons of \cite{CM}.  The fundamental string
Page charge of this tube is one unit.
However, the D2-brane itself
necessarily has two units of D2-brane charge.  Multiple holomorphic curves lead
to an integer number of fundamental strings and an even number of D2-branes.
Thus, the T-dual D0/D8 system necessarily contains an even number of D0-branes.
This argues for the $Z_2$ confinement effect described above.
We close with a discussion of various issues in section \ref{Disc}, including
extensions to non-BPS configurations and a comparison with similar $Z_2$
confinement
effects for five-branes in type I string theory \cite{EWI,GP}.

\section{Half-branes and the Reduction of Holomorphic Curves}
\label{jred}

We consider below the holomorphic curves of \cite{BG} in
the Kaluza-Klein monopole background and reduce
them to configurations of test D2-branes and fundamental strings
in the background of a unit charged D6-brane.
Supposing that the D6-brane is oriented along the $x_0,x_1,x_2...x_6$
directions and introducing $dx_{\parallel}^2 = -dx_0^2 + dx_1^2
+ dx_2^2 + dx_3^2 + dx_4^2 + dx_5^2 + dx_6^2$ and
$dx_\perp^2 = dx_7^2 + dx_8^2 + dx_9^2$, the type IIA background fields take
the form

\begin{eqnarray}
\label{10Dsol}
ds^2_{string} &=&  V^{-1/2} dx_\parallel^2 + V^{1/2} dx_\perp^2, \cr
e^{2\phi} &=& V^{-3/2}, \cr
A_1 &=&  \frac{1}{2} (1 - \cos \theta) d \psi, \cr
F_2 &=& \frac{1}{2} \sin \theta  d \theta \wedge d\psi
\end{eqnarray}
where $V = 1 + \frac{1}{2r}$, $r =x_7^2 + x_8^2 + x_9^2$,
$\theta = \cos^{-1} \left( \frac{-x_9}{r} \right)$,
and $\psi = \tan^{-1}\left( \frac{x_8}{x_7} \right)$.
Here, to simplify the formulas we have set the radius $R_{10}$
of the M-theory circle to one.

We will be interested in solutions in which the D2-brane is extended in
the $x_7,x_8$ directions.  As a result, one can visualize this system
by suppressing the dimensions along the D6-brane and drawing only
the three space $x_7,x_8,x_9$, in which the D6-brane appears as a point
object.   Here, we consider the case where the branes
do not intersect, but instead the D2-brane will pass above
or below the D6-brane.  This is analogous to the $\psi_0 = \pi$ case of 
\cite{Pelc}.  The intersecting case will be investigated in section
3.

\begin{center}
\includegraphics{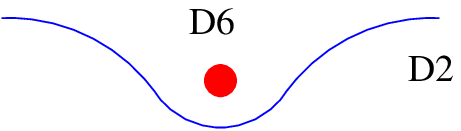}

{Fig. 1.  A sketch of our brane configuration.}
\end{center}

\subsection{The Brane Source Charge}

It is convenient to describe the D2 brane by the electric current it
carries.  When Chern-Simons terms are relevant, there are several
different definitions of a charge current, each of which is useful in
certain settings.  For definiteness, we consider the M2 `brane-source' current
$j_{M2}^{bs}$
of \cite{3Q}, which is the current that arises by varying the action with
respect
to the three-form gauge field $A^M_3$ of eleven-dimensional supergravity
and multiplying by $2\kappa_{11}^2$, the gravitational coupling in eleven
dimensions. Our normalizations are fixed by stating that the bosonic
part of the eleven-dimensional supergravity action is
\begin{equation}
S_{\rm bosonic} = \frac{1} {2 \kappa_{11}^2 } \int d^{11} x (-g)^{1/2}
\bigl( R - \frac{1}{2} |F^M_4|^2 \bigr) - \frac{1}{6}
\int A^M_3 \wedge F^M_4 \wedge F^M_4,    
\end{equation}
where $F_4^M = dA_3^M.$
Note that, when the Chern-Simons term above can be neglected, the 
equation of motion for $A_3^M$ in the presence of a source is 
\begin{equation}
d * F_4^M = *j^{bs}_{M2}. 
\end{equation}
One can check that, since no M5-branes are present, the supergravity equations
of motion imply conservation of this current $d*j^{bs}_{M2}=0,$ and
that the brane-source current agrees with the Maxwell and Page
currents \cite{3Q}.

We will be interested in calculating the D2-brane and fundamental
string (F1) brane-source currents resulting from the Kaluza-Klein reduction to
ten dimensions.   Thus we consider eleven-dimensional solutions with
a spacelike Killing vector field $\lambda^\alpha$, normalized so that
$|\lambda| =1$ at infinity.
Again the F1 and D2 brane-source currents in type IIA are
given by varying the action with respect to gauge fields, in this case
the Neveu-Schwarz two-form $B_2$ and the Ramond-Ramond 3-form $A_3.$
A useful way to write the relation between these currents and $j_{M2}$
is in terms of the
one-form $\tilde \lambda$ constructed from the Killing field
$\lambda^\alpha$ by lowering the index and renormalizing to
set $\tilde \lambda_\alpha \lambda^\alpha =1;$ i.e., we set
$\tilde \lambda_\alpha = \frac{\lambda_\alpha
}{|\lambda|^2}.$
Such a one form can be expressed in terms of a coordinate $x_{10}$ on the
Killing orbits
and the associated Kaluza-Klein gauge field $A_1$ as $\tilde \lambda = dx_{10}
+A_1.$
One can then verify from the supergravity equations of motion that the currents
are related by

\begin{equation}
\label{j2red}
*j^{bs}_{M2} = *_s j^{bs}_{D2} \wedge \tilde \lambda - *_s j^{bs}_{F1},  \ \
\ \rm{or} 
\end{equation}
\begin{equation}
\label{2j2red}
j^{bs}_{M2} = e^{2 \phi/3} j^{bs}_{D2} - e^{8\phi/3} j^{bs}_{F1} \wedge
\tilde \lambda.
\end{equation}
Here, the star ($*$) denotes the Hodge dual with respect to the metric of
11-dimensional supergravity, while the star ($*_s$) with an $s$ subscript 
denotes the Hodge dual with respect to the type IIA string metric.
The usual type IIA dilaton is represented by $\phi$.  The ten-dimensional
spacetime ${\cal M}_{10}$ is the quotient of the 11-dimensional
spacetime ${\cal M}_{11}$ by the $S^1$ orbits of $\lambda^\alpha$
and the type IIA currents have
been pulled back to the 11-dimensional spacetime via this quotient map.  
Equation (\ref{j2red},\ref{2j2red}) is
one precise version of the statement that fundamental strings are associated
with that part of the M2-brane current that flows around the $x_{10}$
direction.

A useful form of the metric for the 11-dimensional
Kaluza-Klein monopole is
\cite{NOYY,Yosh,BG}

\begin{equation}
ds^2 = - dx_\parallel^2 + V dv d\overline v + V^{-1} \left|
\frac{dw}{w} - f dv \right|^2,
\end{equation}
where
\begin{equation}
f = \frac{x_9 + r}{2vr},
\end{equation}
correcting a small typographic error in \cite{NOYY,BG}.
The complex coordinates
$v$ and $w$ define one of the complex
structures on the Euclidean Taub-Nut space.  They are related to the
ten-dimensional coordinates through
\begin{eqnarray}
v &=& x_7 + i x_8 \cr
w &=& e^{-(x_9+ix_{10})} \left( - x_9 + \sqrt{x_9^2 + |v|^2} \right)^{1/2},
\end{eqnarray}
and Kaluza-Klein reduction takes place along the Killing  field
$\partial_{x_{10}}$.
Such coordinates are smooth so long as $v \neq 0$ or $x_9 < 0$.
Note that $x_{10}$ ranges over $[0,2\pi]$ consistent with our
setting $R_{10}=1$.

Reference \cite{BG} in fact describes
two interesting holomorphic curves given by $w =  e^{-b}$ and
$w = e^{-b} v$ for some complex constant $b$.
A symmetry of the form $(w,v) \rightarrow (\frac{v}{w}, v)$
interchanges these two families of curves, so that we need only
consider one of them.  A useful observation
is that this symmetry changes the sign of $x_9$, but leaves $r$
invariant.  When
$b + \overline b$ is large and
negative, the first curve describes a mostly flat M2-brane located
at large negative $x_9$ oriented along the $x_7$, $x_8$ directions.
For large positive $b + \overline b$,
this curve is tightly cupped around
the monopole.

Let us consider the first curve, $w= e^{-b}$, in detail.
One can check that this curve has $x_9 < 0$ for $x_7 = x_8 = 0$ and so
is manifestly smooth.
Some calculation shows that an M2-brane lying on such a curve at
$x_1,x_2,x_3,x_4,x_5,x_6=0$ is associated
with the current
\begin{equation}
j^{bs}_{M2} = \frac{i2 \kappa_{11}^2T_{M2}}{2}
\delta^{(6)}(x) \delta^{(2)}(w - e^{-b}) \frac{w \overline w}{V^2}
dx_0 \wedge \omega \wedge \overline \omega,
\end{equation}
where
\begin{equation}
\omega = ( V^2 + f \overline f) dv - \overline f \frac{dw}{w},
\end{equation}
and $T_{M2}$ is the tension of the M2-brane.

To pick out the part of $j_{M2}^{bs}$ associated with fundamental strings, we
see from (\ref{2j2red}) that
one need only contract $j^{bs}_{M2}$ (on the final index) with the Killing
field
$\lambda =  \partial_{x_{10}}$.    At this stage, it is useful to
introduce the cylindrical radial coordinate $\rho = |v|$.
The result takes the form
\begin{equation}
\label{almostF1}
j^{bs}_{M2} \cdot \lambda = -  \frac{2 \kappa_{11}^2T_{M2}}{2} 
\delta^{(6)}(x) \delta^{(2)} (w -
 e^{-b})
\frac{ \tilde f w \overline w}{\rho} dx_0 \wedge(d \rho + \frac{\tilde f}{2V
\rho} dx_9),
\end{equation}
where we have introduced $\tilde f = 1 - \cos \theta$ so that the type IIA
magnetic 1-form potential is
$A_1 = \frac{1}{2} \tilde f d \psi$. One sees
that (\ref{almostF1}) does not project directly to the ten-dimensional
spacetime as the delta functions depend on $x_{10}$.  This is merely
a reflection of the fact that translations along the Killing field
$\lambda$ do not map our holomorphic curve onto itself.  Nevertheless, at events
that are far from the curve as compared with the size of the $S^1$ orbits,
type IIA supergravity should still be a good description
of the spacetime obtained
by placing an M2-brane on this curve.  In this regime, the type IIA
fields are related to the average values of the current over the $S^1$.
Let us therefore
average the current (\ref{almostF1}) over the orbits of the Killing field by
multiplying by $1/{2\pi}$ and integrating over $dx_{10}$.  
Multiplying
(\ref{almostF1}) by the appropriate dilaton factors 
and adjusting the sign as dictated by
(\ref{2j2red}) yields the associated fundamental string current.  In
the curved spacetime context, it is the dual current $*j_{F1}$
which is most easily interpreted.  Here, $*$ represents the Hodge
dual defined by the string metric of the D6-brane spacetime.  The
result may be written

\begin{equation}
\label{F1cur}
* j_{F1} = \frac{\tilde f 2 \kappa_{10} T_{F1} }{2}
\delta^{(6)}(x)  \wedge_{i=1}^6 dx_i \wedge
\left(  \delta( x_9 - \hat{x}_{9}(\rho))
 \frac{d \psi}{2\pi} \wedge dx_9
 + \delta(\rho - \hat{\rho}(x_9))
d \rho \wedge \frac{d \psi}{2\pi} \right),
\end{equation}
where $\hat{x}_{9}(\rho)$ and $\hat{\rho}(x_9)$ are the functions determined
by the relation $|w| = |e^{-b}|,$ the factor
$T_{F1}$ is the tension of
a fundamental string, and $2\kappa_{10}$ is the ten-dimensional gravitational
coupling.  In arriving at (\ref{F1cur}) we have used the relation
$2 \kappa_{11}^2 T_{M2} = 2\kappa_{10}^2 T_{F1}.$

Note that, due to the smearing, the fundamental string
current $j_{F1}$ in fact has support on
a submanifold with two spatial dimensions.  This submanifold
is just the worldsheet of the D2-brane obtained by dimensional reduction
of the holomorphic curve.  Thus, the strings are dissolved in the D2-brane, and
the string charge runs outward along the D2-brane in the radial direction.
Note in particular that this configuration is
rotationally symmetric about the $x_9$ axis.   The resulting shape
of the D2-brane is quite similar to that of \cite{Ima,CGS,GM}.

In the above form, the number of fundamental strings passing through
a hypersurface $x_0 = constant$ with either $\rho = const$ or $x_9 = constant$
is readily seen to be $\frac{\tilde f}{2} = \frac{1 - \cos \theta}{2}$,
where $\theta$ is determined by the intersection of the hypersurface
with the worldsheet of the M2-brane (or, equivalently, of the resulting
D2-brane).  The fundamental string current vanishes
on the $x_9$-axis, but represents half a string exiting
to infinity (where $\theta = \pi/2$ due to the fact that
the D2-brane becomes very flat).

It is instructive to note what occurs when the parameter $b$ is adjusted so
that
the D2-brane appears to move far past the D6-brane.  In this case, the
surface of the D2-brane deforms as shown below.  One can see that,
at intermediate values of $x_9$, the surface takes the shape of a long
thin tube; i.e., a string.  In such a configuration, the gauge field
strength $\tilde F_{[4]}$ that couples to D2-branes is nearly zero,
as a cylinder has no net D2-brane charge.  Instead, the string
charge will dominate the coupling to the supergravity fields\footnote{Some
of the details can be found in the appendix of \cite{GKMT} for the 
more familiar BIons of \cite{CM}.}.
On this string, $\theta$ is essentially zero so that the string
carries a full unit of fundamental string brane-source charge.

\begin{center}
\includegraphics{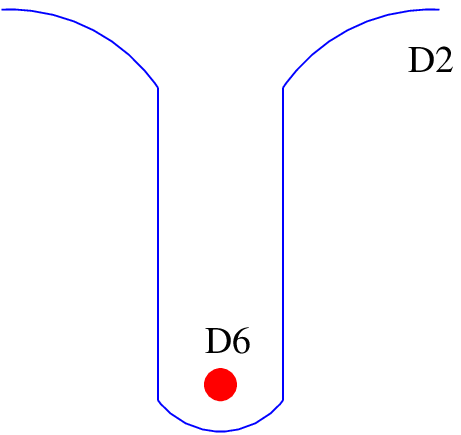}

{Fig. 2.  A D2-brane is moved far past a D6-brane.}
\end{center}

To understand the general features of the above configurations, we recall from
\cite{3Q} that the brane source currents are not in general conserved.
In particular, taking an exterior derivative of (\ref{j2red}) and using
conservation of
$*j_{M2}^{bs}$ (which holds in the present context free of M5-branes) yields
the relation
\begin{equation}
\label{F1C}
d*j^{bs}_{F1}  = - * j_{D2}^{bs} \wedge F_2.
\end{equation}
We refer to this equation as a constraint as it has components
that involve no time derivatives.  Here, $F_2 = dA_1$ represents the magnetic
flux from the D6-brane.  The normalization is such that, as our D2 brane
will capture half of the flux from the D6-brane, it is in fact
a net source of one-half unit of brane source charge.  To see this, note
that the total flux (\ref{10Dsol}) from the D6-brane is
$2\pi R_{10} = 2\pi$, which matches the ratio $\frac{T_{F1}}{T_{D2}}$.
An analogous
discussion from the point of view of the worldvolume theory can be
found in \cite{DFK,K}.
Such string charge is created in our solution though the resulting
`half-string' does not leave the D2-brane.  Instead, it
flows out along the D2-brane to infinity.   The $\tilde f$ factor in
(\ref{F1cur}) describes the gradual creation of this 
fundamental string current as $F_2$ flux is captured by the
D2-brane.  From this and the normalization of (\ref{F1C}) we can verify
that, as expected, the associated $j^{bs}_{D2}$
represents exactly one D2-brane lying on the surface $\rho = \hat{\rho}(x_9).$

\begin{center}
\includegraphics{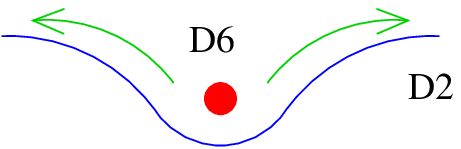}

{Fig. 3.  The D2-brane carries F1 charge radially outward.}
\end{center}

One might ask if the D6-brane should also be the source of a string.  Indeed,
there
is a constraint in type IIA theory which states that a D6-brane should be the
source
of fundamental string brane-source current when it captures an electric flux of
the gauge field
$\tilde F_4$ associated with a D2-brane.  Here, the relevant quantity is the
`improved
field strength' $\tilde F_4 =  dA_3 - A_1 \wedge H_3$, which is the fully gauge
invariant
quantity.  The electric flux is computed in terms of the dual $*_s\tilde F_4.$
Now, since we have treated the D2-brane as a test object, the associated $A_3$
gauge
flux is not readily apparent.  Nevertheless, a short argument shows that the
associated
field strength $*\tilde F_4$ must in fact vanish at the D6-brane.  The point is that
this field strength
is related to the 4-form $F_4^M$ of 11-dimensional supergravity through
\begin{equation}
*F_4^M = - e^{-2\phi} *_s H_3 + *_s \tilde F_4 \wedge \tilde \lambda.
\end{equation}
Now, since the M2-brane does not intersect the core of the Kaluza-Klein
monopole,
$*F_4^M$ should be smooth there.  Since $\lambda$ is a smooth vector field, the
contraction $*F_4^M \cdot \lambda$ must be smooth and must vanish whenever
$\lambda$
vanishes, such as at the core of the monopole.  Since this contraction is just
$* \tilde F_4$, we see that see that this flux vanishes at the D6-brane.

\subsection{The Maxwell and Page Charges}

Let us now look at the Maxwell current defined \cite{3Q} by $d \left( e^{-2\phi}
*_sH_3 \right) = *_sj_{F1}^{Maxwell}.$
This may be written in terms of the brane source current as
\begin{equation}
\label{eqREF}
*_sj_{F1}^{Maxwell} = *_sj_{F1} + *_s \tilde F_4 \wedge F_2.
\end{equation}
Maxwell charge is always `diffuse,' in the sense that it is carried by the bulk
fields and so
is not localized in branes.  Some useful insight into the distribution of
Maxwell charge in this
system can be obtained by integrating the second (diffuse) term over
conveniently chosen volumes.  In particular,
suppose that we integrate this term over an 8-surface $V_8$ defined by $\rho =
\rho_0$ for
$ x_9^{min} < x_9 < x_9^{max}$ and $\psi^{min} < \psi < \psi^{max}$.  The
coordinates $x_1,...,x_6$
are allowed to run from $-\infty$ to $\infty.$

In the approximation that the D2-brane is a test object,
$F_2$ is independent of $x_1,...,x_6$.  Thus, we may begin by integrating
$*_s \tilde F_4$ over each surface $x_9 = x_9^0, \rho = \rho_0, \psi = \psi_0.$
The result must be one-half the charge of the D2-brane
times a sign, depending on whether this surface passes above or below the
D2-brane\footnote{In
the present setting without D4 or NS5-branes, the three notions of D2-brane
charge from \cite{3Q}
all agree.}.
As a result, the diffuse charge in $V_8$ is 1/2 of the flux $F_2$ captured by
$V_8$ above the
D2-brane minus 1/2 of the flux captured below the D2-brane.
In particular, for $\psi^{min} = 0, \psi^{max} =2 \pi$, we have
\begin{equation}
Q_{F_1,V_8}^{Maxwell} = \int_{V_8} *_s j_{D2}^{Maxwell} = \frac{T_{F1}}{2}
\left( 1 - \cos \theta_0 \ \  {+ F_2  \ {{\rm flux \  above \ brane}} \atop { -
F_2  \ {\rm flux \  below \ brane}}} \right),
\end{equation}
where $\theta_0$ is the value of $\theta$ at which $V_8$ intersects the
D2-brane.

Note that, as mentioned in \cite{3Q} for the similar D4/D6 case,  the
Maxwell charge remains diffuse for large $\rho_0$; i.e., `near infinity.'
Even when $V_8$ is
contractible, $Q_{F1,V_8}^{Maxwell}$ need not vanish and, by
an appropriate choice of $V_8$, $Q_{F1,V_8}^{Maxwell}$ can be made to take
on a continuum of values.  Perhaps the most interesting case is when
$V_8$ is taken to be the entire cylinder (or, equivalently, the entire sphere)
at infinity.  This might be said to give the total Maxwell string charge created
in the spacetime and flowing to infinity.  Since $\theta_0 \rightarrow \pi/2$
for large
$\rho_0$, the $F_2$ fluxes above and below the brane cancel and the total
Maxwell
charge flowing to infinity is just $T_{F1}/2.$  In this case, Maxwell
charge does not shed any particular light on charge quantization.  The
difference
between the present setting 
and that of \cite{Taylor} can be traced to the presence
of diffuse charge at infinity.

Finally, let us consider the Page current defined by
\begin{equation}
\label{PageF1}
*_s j_{F1}^{Page} = d(e^{-2\phi} *_sH_3 - A_1 \wedge *_s \tilde F_4)
= *_s j_{F1}^{bs} - A_1 \wedge *_s j_{D2}^{bs}.
\end{equation}
Although this current is gauge dependent, the corresponding charge is
naturally quantized.  Integrating (\ref{PageF1}) over some $V_8$ yields a
charge
which is invariant under small diffeomorphisms, though it transforms under
large
diffeomorphisms.  A study of the Kaluza-Klein reduction from 11-dimensions
shows
that the F1-brane Page charge associated with some region $V_8$ is identical
to the M2-brane charge in the 8-volume given by lifting $V_8$ to the
11-dimensional
spacetime using the supplementary condition $x_{10}= constant.$  The change
of the Page charge under a large diffeomorphism $A_1 \rightarrow A_1 - d
\Lambda$
just corresponds to the fact that the $x_{10}$ coordinate transforms as
$x_{10} \rightarrow x_{10} - \Lambda$, so that $V_8$ may now lift to a
different surface in 11-dimensions.  Since the Page charge is quantized, it
should
not lead to the discovery of any half-branes.

Let us now work out the Page charge for our solution.   It is clear from
(\ref{PageF1}) that Page charge will be localized on the brane.
When dealing with the Page charge, it is important to avoid
integrating through a Dirac string.  We therefore choose the
gauge $A_1 = \frac{1}{2}(1  - \cos \theta)d\psi$ with the Dirac string along
the positive $x_9$ axis.  
Consider then the surface
$V_8$ described just after eq. (\ref{eqREF})
for $\psi^{min}=0,\psi^{max}=2 \pi.$
If this surface is not to intersect the D6-brane, we should
have $x_9^{min},x_9^{max} < 0.$  In this case, $V_8$ has
two boundaries $\Sigma_{min}, \Sigma_{max}$ at
$\rho = \rho_0$, $x_9 = x_9^{max}, x_9^{min}.$  Thus, we have
\begin{equation}
\int_{V_8} *_sj^{Page}_{F1} = \left(\int_{\Sigma^{max}} -
\int_{\Sigma^{min}} \right) (e^{-2\phi} *_s H_3 + *_s \tilde F_4 \wedge A_1).
\end{equation}
Now, the supergravity equations of motion state that
\begin{equation}
d ( e^{-2\phi} *_s H_3 - *_s \tilde F_4 \wedge A_1) = 0
\end{equation}
when the brane-source charge vanishes.  Thus, the integrals over
$\Sigma^{min}, \Sigma^{max}$ are unchanged by a homotopy that
does not move them through any branes or Dirac strings.
But, in our case, both surfaces are contractible.  Thus,
both integrals vanish and the Page charge is zero.

It is also interesting to compute the Page charge in other gauges
of the form $A_1 = \frac{1}{2}(1 + 2n - \cos \theta) d \psi$ for integer $n$.
Under a change of gauge, we see that the Page current changes
by $\Delta j^{Page}_{F1} = -  \Delta A_1 \wedge *_s j_{D2}^{bs}.$
Thus, in the gauge labelled by $n$, there are $n$ units of fundamental string
Page
charge flowing outward along the brane to infinity.  As expected, the Page
charge
is quantized.  One can think of this charge as entering the D2-brane along the
Dirac string.  From the above discussion of Kaluza-Klein reduction, it is clear
that it is the Page charge that appears in the supersymmetry algebra and indeed
this is the standard result (see , e.g., \cite{Stelle}).
It is therefore surprising that one can find fundamental
string charge radiating to infinity in what, by construction, is a solution
with
Killing spinors.  Certainly, in the absence of the 6-brane, a bundle of
fundamental
strings that fan radially outward would be sure to break supersymmetry.
It seems that the complicated asymptotic structure of our solution
makes supersymmetry subtle in gauges where the D2-brane is
topologically non-trivial due to its intersection with the Dirac string.

In the present setting, we have seen that the issue of half-branes is
really a question of proper accounting.  The constraint requires
that the D2-brane be the source of 1/2 unit of fundamental string
brane source charge.  This is not a problem as in general it is the
Page charge that should be quantized and not the brane source charge.
Indeed, we have seen that the Page charge takes integer values.

However, in the above example a half
unit of brane source charge does flow to infinity along a nearly flat
D2-brane.  Note that since brane source charge is a physical
(i.e., gauge invariant) notion,  such a brane source current
will prevent compactification of the D2-brane, despite the vanishing of
the Page charge in the simplest gauge.  The issue is similar to the familiar
fact that the total electric charge must vanish on a compact space.
Thus, the example considered here is not related by T-duality to the D0/D8
case.

In order to be compactified, a mostly flat D2-brane 
must have an induced F1 brane-source current
that vanishes at infinity.  However, considering a spacetime of the form
${\bf R}^8 \times T^2$ in which a D2-brane
wraps some $T^2$ transverse to a D6-brane, we see that such a D2-brane
must still be the source of a net 1/2 unit of string brane-source charge.
Thus, the D2-brane must deform so that either 1) a part of it represents
a string running off to infinity in some remaining non-compact direction
or so that 2) it intersects the D6-brane, into
which the F1-brane charge can flow. In either case, mere accounting
issues cannot rid us of the factor of 1/2.  To see this, we remind
the reader that the brane source and Page currents for fundamental
strings differ only by a term of the form $A_1 \wedge *_s j_{D2}^{bs}.$
In a region where the D2-brane takes the shape of a long narrow tube (i.e.,
a string), we can arrange to move the Dirac string elsewhere and to have
$A_1$ smooth near the string.  If the tube is thin\footnote{One might
also consider configurations in which a finite width cylinder of D2-brane
extends to infinity.  Non-BPS initial data of this sort
can be given in certain cases, but energy arguments imply that
this cylinder should contract.}, integrating $A_1$
around the tube must then give zero.  Thus, under such circumstances, the
brane source and Page charges of the string will be equal.
If the D2- and D6-branes meet and intersect, we can create such
a thin tube by moving the flat part of the D2-brane far from the D6-brane
(in analogy with Figure 2).

We conclude that an understanding of cases in which the worldvolume of
the D2-brane is compact  requires an analysis of the actual
intersection of the D2- and D6-branes.  Due to the D6-brane
singularity,
it is not clear from the type IIA perspective which intersections are in fact
allowed.  Since, however, the core of a unit charged Kaluza-Klein monopole is
smooth,
the question is much more tractable in the 11-dimensional formulation.
It is to this question that we turn in the next section.

\section{Intersecting D2- and D6-Branes}
\label{hol}

We have seen that, starting with a smooth holomorphic
curve in the Kaluza-Klein monopole spacetime, dimensional reduction can
yield a type IIA configuration in which 1/2 unit of fundamental string
charge exits to infinity.  In this case, the fundamental string charge
is dissolved within a D2-brane.  One might wonder whether one can
in  fact find holomorphic curves whose reduction to ten dimensions
yields 1/2 strings in more familiar string-like configurations which
are not dissolved inside the D2-branes.  Here, one may take inspiration from
\cite{CGS}, where a number of solutions were found
to the Born-Infeld equations of motion
for a D5-brane in the background of a D3-brane which yield string-like
configurations when parameters are properly adjusted. Similar configurations
for D3-branes in a background of NS5-branes were recently constructed
in \cite{Pelc}.

The solutions studied in \cite{CGS,Pelc} were rotationally symmetric and
BPS\footnote{The authors of \cite{CGS} argued that their solutions were likely
to be BPS, while the solutions of \cite{Pelc} are conclusively so.}.  
When the test D-brane was moved far past the background
brane, the world volume of the test brane is deformed such that it
forms a string stretching between the main part of the test brane and
the background brane.
The tension of this string was computed in \cite{CGS} and was found
to depend on the total amount of background flux that had been captured
by the test-brane.  Although the context is slightly different, this is
in direct parallel with the constraint (\ref{F1C})
which describes the creation of strings.
Solutions that appeared to contain fractional strings
arose when the test D-brane intersected the background brane.
The configuration is illustrated in the diagram below.
The test brane formed a cone near
the background brane, and the tension of the string was related to the
opening angle of this cone.    In the symmetric case where the opening
angle is $\pi/2$, the tension was found to vanish.  In \cite{Pelc}, it
was argued that a corresponding opening angle was quantized for consistency
with charge quantization.

\begin{center}
\includegraphics{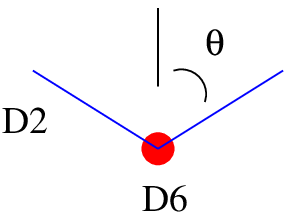}

Fig. 4.  The D2-brane takes the shape of a
cone with opening angle $\theta$ and vertex at the D6-brane.
\end{center}

The authors of \cite{CGS} question the physical relevance of such
solutions, as the Born-Infeld description of the test-brane
cannot describe the details of the intersection between the test-brane
and the background brane.  It is not clear that this should have been a
concern,
as the background generated by a highly charged D3-brane is a large
black brane with a smooth horizon, which is well described by
supergravity.   In any case, the situation here is well under control.
When lifted to M-theory,
the core of the D6-brane becomes just the center of a unit charged
Kaluza-Klein monopole, which is perfectly smooth.  As a result, precisely
those solutions which correspond to a smooth M2 test-brane
should reduce to a valid test brane configurations in type IIA.

In order to find such a solution, let us note that the simplest
test-brane configuration with opening angle $\pi/2$ is simply given
by placing the D2-brane on the surface $x_9=0$.   A holomorphic
curve that reduces to this surface can be found by recalling that
the symmetry $(w,v) \rightarrow (\frac{v}{w},v)$ changes the sign
of $x_9$, so that any surface which is invariant under this symmetry
must lie at $x_9=0$.  One can readily verify that the surface
$w^2 = v$ is in fact invariant.  However, because it contains $w^2$,
upon dimensional reduction we find {\it two} D2-branes lying at $x_9=0$.
The corresponding sheets of the M2-brane lie at $x_{10} = \psi$ and
$x_{10} = \psi + \pi.$
Deforming this surface to move the asymptotic parts of the D2 branes
to large $|x_9|$ therefore produces a string-like piece of D2-brane
carrying a full unit of fundamental string charge.  Note that
the two D2-branes cannot be separated from one another as, when
the angle $\psi$ increases by $2\pi$, we must move from one brane
to the other.  Note also that a ${\bf Z}_2$ quotient of this space leads
to the charge 2 Kaluza-Klein monopole and the charge 2 D6-brane.
Such a ${\bf Z}_2$ quotient identifies the two sheets of the M2-brane, leading
to a single D2-brane in the charge 2 D6-brane background.

For completeness,
let us consider the most general holomorphic curves that lead to rotationally
symmetric configurations in type IIA theory. That is to say, we wish
to consider curves such that if the curve intersects $m$ times with
the fiber over
$(v, x_9)$, then it also intersects $m$ times with
the fiber over $(e^{i \psi_0} v, x_9)$.
This implies that the holomorphic curve is of the form
$v^n = e^b w^m.$   Since $v,w$ are smooth coordinates away
from the positive $x_9$ axis (i.e., $\rho =0$, $x_9  \ge 0$),
any $v^n = e^b w^m$ that does not
intersect this axis is in fact smooth. Similarly, the $Z_2$ symmetry
tells us that any such curve which avoids the negative $x_9$
axis ($\rho=0, x_9 \le 0$) is smooth.
As a result, we may conclude that all curves
$v^n = e^b w^m$ which do not intersect the origin $\rho = x_9 = 0$
are smooth.

To study the curves near the origin, we note
that $x_0,x_1,...x_6$ together with the complex
coordinates $\beta = w, \alpha = v/w$ form a smooth coordinate
system near $x_7 = x_8 = x_9 =0$.  One may check that near this point
the metric takes the form
$ds^2 = d\alpha d\overline{\alpha} + d \beta d \overline{\beta} + O(\alpha)
+ O(\beta).$
Note that the $Z_2$ symmetry merely interchanges $\alpha$ and $\beta$.
It is useful to note here that our $\alpha$ is the complex coordinate
called $v$ in \cite{GM, Aki} while our $\beta$ is the complex coordinate
called $\overline w$ there.
Since the curves may be written $\alpha^n = e^b\beta^{m-n}$, it is
clear that curves intersecting the origin
are smooth for $n=1$ (and $m-n>0$)
or $n = m-1 > 0$, but not for other cases.
The $Z_2$ symmetry interchanges $n=1$ with $n = m-1$.

What is interesting about all of this is that, from the point of view
of the IIA spacetime, the curves do not appear to intersect the D6-brane
in a smooth way.  The curves satisfy
\begin{equation}
\label{curve}
x_9 = \frac{1}{2} e^{\frac{b + \overline b}{m}}  e^{-2 x_9} \rho^{2(1 - n/m)}
-  \frac{1}{2} e^{-\frac{b + \overline b}{m}}  e^{+2 x_9} \rho^{2n/m},
\end{equation}
so that near the origin they will in general have the form of a cusp.
Let us take the case $n=1$.  Such curves intersect the origin smoothly only
for $m \ge 2$, in which case we find:
\begin{equation}
x_9 \sim \rho^{2n/m}.
\end{equation}
For the special case $n=1$, $m=2$ which we argued might give rise to
half-branes, the curve is a cone near the origin, satisfying
\begin{equation}
x_9 \sim \rho \sinh(\frac{b + \overline b}{2}).
\end{equation}
Since these results correspond to smooth holomorphic curves in the
eleven dimensional space, we conclude that they do in fact describe
BPS configurations of D2 branes intersecting a D6-brane.  

It is interesting to compare our cone-shaped solutions ($n=1,m=2$)
to those of \cite{CGS}.  Our solutions are more restrictive, as
the asymptotic (large $\rho$) location of our D2-branes
is correlated with the opening angle of the cone.   In particular, for
large $\rho$ our $n=1,m=2$ solutions
approach\footnote{As may be verified from (\ref{curve}),
the usual logarithmic divergence associated with
a string intersecting a D2-brane is absent in this case.}
$x_9 = \frac{b + \overline b}{4}$.
In contrast, the opening angle and asymptotic location
were separate parameters in \cite{CGS}.   This might possibly
be due to the lower (co-)dimensional 
nature of our system.  For fundamental strings
attached to a larger brane, the brane should flatten more quickly
at large $\rho$ which might allow the asymptotic $x_9$ position and the
opening angle to be tuned separately.

We should, of course, evaluate the flux of fundamental strings carried
by the general curve,
$w^m = e^{-b} v^n.$  The calculation proceeds much as in section
\ref{jred}.  The result is that the flux of strings
passing through a surface $\rho=constant$ or $x_9=constant$ is given by:
\begin{equation}
\label{flux}
{\rm Brane-source \ F1 \  Flux}  = \frac{1}{2} [(m-2n) - m \cos \theta],
\end{equation}
in units of fundamental string charge quanta.
At large $\rho$, one may check that the surface satisfies
$x_9 = \pm \left| \frac{1}{2} - \frac{n}{m} \right| \ln \rho$
so that, in particular, $\theta \rightarrow
\pi/2$ as $\rho \rightarrow \infty.$  Thus, it follows 
that such configurations
have net flux of $\frac{m-2n}{2}$ strings running out to infinity (and not
those flowing into the D6-brane)
along $m$ D2-branes.  We note that it is the number of strings exiting
to infinity that controls the asymptotic shape of the D2-brane.
This observation may be important for fixing the proper relationship
between the tension
computations of \cite{CGS} and charge quantization.
The coefficient of $\ln \rho$ is just what one would find for $m$ D2-branes
attached to $\frac{m-2n}{2}$ fundamental strings.

These configurations contain $m$ D2-branes, so that the term
$\frac{m}{2}\cos \theta$ is consistent with our constraints determining the rate
of creation of string charge.  Note that in regions where $\theta =0$
and $\cos \theta = 1$, the total flux of strings is always an integer ($n$).
Due to the symmetries of our setting, long thin tubes approximating strings
can arise only at such values of $\theta$.  In particular, for $n=1, m>2$, 
the tube always becomes thin at the cusp where it intersects the
D6-brane.  Thus, an integer amount of Page charge flows into the D6-brane and
the D2- and D6-branes may be said to be connected by an integer number of
branes.

The most interesting case is $n=1,m=2$ 
for which the brane source charge
vanishes at infinity.  Note that this case maps to itself (with
$b \rightarrow -b$) under the ${\bf Z}_2$ reflection.
It is only this case that one might hope to
compactify and relate to the D0/D8 system via T-duality. 
As we have seen, the corresponding D2-brane intersects the D6-brane and 
does not run off to infinity.
In the usual way, the Page current is conserved and is related
to the brane-source current by
\begin{equation}
*_s j_{F1}^{Page} = *_s j_{F1}^{bs} - A_1 \wedge *_s j_{D2}^{bs}.
\end{equation}
Conservation means that it suffices to compute the charge
at infinity.
The case $n=1,m=2$ contains two D2-branes, we see that in the
simplest gauge $A_1 = \frac{1}{2}(1 - \cos \theta)$ we have one
unit of Page charge on the brane.  In particular, the string
connecting the D2- and D6-branes has one unit of Page charge.
In the more general gauge
$A_1 = \frac{1}{2}(1 + 2n - \cos \theta)$, we have
$(2n+1)$ units of Page charge.  It appears that, for this
case,  there is no gauge in which the Page charge vanishes.
Similar comments hold for the Page charge in the other smooth
cases.  In all cases, diffuse Maxwell charge is present at
infinity.

\section{Discussion}
\label{disc}
\label{Disc}

In the above work, we have studied smooth holomorphic curves
representing test M2-branes
in the background of a Kaluza-Klein monopole.  The associated
distributions of charge in type IIA theory have been computed for
all cases where the reduction to type IIA is rotationally symmetric.
This has allowed us to construct a number of solutions where the
intersection of the D2-brane and the D6-brane appears singular from
the type IIA perspective, although the 11-dimensional perspective
reveals that it is in fact smooth and BPS.  As a result, we expect
that all of the solutions\footnote{Or, at least, that
dense set for which the string charges can be made integral by considering
some integer number of copies.}
of \cite{CGS} do in fact correspond to valid BPS solutions.

Perhaps the most interesting part of this work is that it
provides insight into issues involving half-strings that arise
in the context of the Hanany-Witten effect.   When a single
D2-brane is stretched above a D6-brane as in Fig. 1
the constraint (\ref{F1C}) requires it to be the source
of 1/2 unit of brane-source charge.  We have seen that in
certain cases where this half-string runs to infinity
along a flat D2-brane, the half-string is merely a matter
of accounting.  In general it is the Page charge, as opposed
to the brane source charge, that is quantized and the Page
charges for such cases turn out to be integral.

However, such a string charge running to infinity along
a mostly flat D2-brane will prevent the brane from being 
compactified.
The effect is much the same as the familiar statement that
a compact space must contain zero total charge.

In the Hanany-Witten setting, one typically assumes
that the brane source charge leaves the D2-brane by
flowing into the larger D6-brane.  In this case, the
D2-brane and D6-brane must intersect.  If we now imagine
pulling the main body of the D2-brane far away from the
D6-brane, they would remain connected by a thin
tube that approximates a fundamental string.
However, we have seen that the brane-source and Page
charges of such a thin tube agree, so that mere accounting
cannot in this case explain the factor of 1/2.

It is therefore natural to suspect, in the case where
the $x_7$ and $x_8$ directions are compactified and the
D2-brane is compact, that only such combinations of branes
can arise for which half-strings are not needed.
In section \ref{hol}, we found some evidence that M-theory 
provides a mechanism to enforce this constraint.
Considering now the uncompactified case,
we succeeded in finding smooth BPS M2-branes in the 11-dimensional
Kaluza-Klein monopole for which a (single) corresponding D2-brane
would be attached to the D6-brane by 1/2 of a fundamental
string.  Furthermore, the F1 brane-source charge for such D2-branes
vanished at infinity so that no obstacle to compactification is expected.
However, in order for our 11-dimensional surfaces to be smooth,
they necessarily projected to a {\it pair} of D2-branes which were in fact
joined to the D6-brane by a single unit-charged fundamental string.
Thus, our results suggest that in a compactified background which contains
a torus ($T^2$) transverse to the D6-brane, compact D2-branes wrapping
this torus can arise only in pairs.  In contrast, we were able to
construct isolated D2-branes in the background of the charge 2
D6-brane -- a case where again only integer charged strings are
required.

The reader may rightly ask whether one can trust such classical
arguments to tell us about fine points of charge quantization.
It is clear that our description
does have a regime of validity.  When the $x_{10}$ circle is large compared
to the eleven-dimensional Planck scale, 
the Kaluza-Klein monopole description of the background and is very
flat on the gravitational scale associated with an M2-brane.
Thus, we expect that a collection of $N$ M2-branes for $N \gg1$
is well-described by a test brane in this background if the $x_{10}$
circle is large enough.
In this case it is clear that the corresponding IIA system has $2N$ units
of D2-brane charge to leading order in $N$.  However, if the actual
D2-brane charge were $2N+1$, so that the number of D2-branes was odd, 
then this would not resolve our dilemma about half-strings.  While it is not
clear how to rigorously rule out such $1/N$ corrections, the final
picture is quite compelling.  One might therefore say that our results
constitute a strong plausibility argument rather than a proof.
It is important to bear this in mind, but we will not consider such
subtleties further here.

Let us take a moment to investigate this confinement mechanism more carefully
by considering the D6-brane background with $x_7$ and $x_8$ compactified, and
by dropping the restriction to the BPS cases.
Equivalently, we may discuss periodic surfaces in
a periodic two-dimensional array of unit charged D6-branes.
Such a solution lifts to a multi-center Kaluza-Klein monopole metric
\cite{multi}
in eleven dimensions.  This metric is again smooth and near the `core' of each
monopole
has just the same structure as the single monopole metric.
We are now interested in the question of classifying periodic smooth 2-surfaces
in this background.  The constraint (\ref{F1C}) guarantees that, if the
surface does not
extend to infinity, it must intersect each of the Kaluza-Klein monopole cores.

Now, near any such intersection the surface must be approximately
described by some plane $a_1 \alpha + a_2 \overline{\alpha}
+ b_1 \beta + b_2 \overline{\beta},$ where $\alpha,\beta$ are
smooth complex coordinates near a given monopole core.  
An orbit of $\lambda = \partial_{x_{10}}$ intersects
such a plane twice.  Thus
we find an even number of sheets of D2-branes.
The sheets in fact form a Riemann surface, as traveling in a circle
about the D6-brane moves us from one sheet to the next.  
Thus, pairs of sheets cannot be globally separated, though
we can always introduce a local
deformation of one sheet to separate it from the other in some
small region of the $x_7,x_8$-plane.  
It is useful to note that an explicit solution which is the analogue
of $v/w^2=1$ in the asymptotically flat case is easy to construct:  this
is again just the plane which is invariant under a ${\bf Z}_2$
symmetry which amounts to a reflection about the plane containing the
D6-branes.

We now comment briefly on a potential exception to our ${\bf Z_2}$
confinement phenomenon.  Consider 
the case where two of the directions transverse to the D6-brane are
compactified on a $T^2$, but the third transverse direction remains
non-compact.  The 
potential loophole would involve a single D2-brane in the
compactified system that does not intersect the D6-brane at all, but
instead extends to infinity in the remaining non-compact direction 
and carries away some amount of fundamental
string brane-source charge.  Such configurations cannot be BPS or even static, 
as without a (half) fundamental string to hold them together, the
D6- and D2-brane repel each other \cite{Lif} and the D2-brane will
retreat to infinity.  Nonetheless, initial data of this sort appears to
exist for the D2-brane Born-Infeld system and it seems to lift smoothly
to eleven dimensions.  It would be interesting to understand that
status of such configurations in detail but, for the moment, we set
aside this possible exception. 

Let us now ask what our results 
have to say about the D0/D8 system.  Intuitively, 
we expect these systems to be related by T-duality.  Unfortunately, it
is difficult to apply T-duality precisely in this case.  The curved
branes make it difficult to apply T-duality in a controlled way in
the perturbative setting \cite{T-dual}, and similarly to apply Buscher
T-duality in supergravity \cite{Bus} we would have to `smear' the system to 
create a Killing symmetry. 

Nevertheless, we expect the D0/D8 system to behave analogously.
This would mean that the symmetric unit charged D8 brane background, 
where the Ramond-Ramond field strength takes the values $\pm1/2$ of the
fundamental quantum on either side, is allowed.  However, in such a background
D0-branes would occur only in pairs.  There is clearly no analogue of the
`local separation' that was allowed for the compact D2-branes.

An interesting question is what gauge group would arise on the
pairs of D0-branes. We recall that the 
type I fivebrane also arises only in pairs, at least as counted in
a sense natural to the type IIB description.  There, the gauge group
on such a pair is SU(2) \cite{EWI,GP}.  In the case of the pair of D0-branes
in the background of a single D8-brane, one again expects the
group to be some projection of U(2); i.e., either SU(2) or U(1)
depending on whether the projection is symmetric or anti-symmetric.  

Now, we do not normally
expect a semiclassical description of the sort used here to tell us
about the detailed structure of the gauge groups.
Also, we realize that in the
D2-brane case the branes are allowed to separate locally and are `confined'
only in some global sense.  Thus, we expect that the gauge group in that
case is locally U(2) but with some global constraint.  To get some
idea about this global constraint, let us consider the case where the
two sheets of D2-brane in fact coincide.  Then, we
can make two observations that suggest that the gauge group in the present
setting should include global U(1) rotations instead of global SU(2) 
rotations.   
First, we recall from section 3 that, although the two D2-branes coincide
the corresponding sheets of the M2-brane are separated in $x_{10}$.     
This is clearly true for the ($n=1,m=2$) case $v/w^2 = e^b.$  Since
the two sheets of the M2-brane do not coincide, we would be surprised
to find a group that is not a subgroup of $U(1) \times U(1).$
Second, although T-duality cannot be used in a rigorous way, we expect
T-duality to map our pair of D2-branes to a pair of D0-branes that
are stuck together, and in particular to a pair of D0-branes located
at the same point in space.  Thus, one expects that holonomies of the 
gauge fields on the D2-branes around the non-trivial cycles of the torus
take values of the form $\left[ {{e^{i \phi}} \atop 0}{0 \atop 
{e^{i \phi}}} \right],$ which do not arise in SU(2).
This is in contrast with the case of type I fivebranes, for which
an M-theoretic explanation of the antisymmetric projection condition leading
to SU(2) can be related to two branes having mirror-symmetric positions
across a would-be orientifold plane.


\acknowledgments

The author would like to thank Phillip Argyres,
Matthias Gaberdiel, Andr\'es Gomberoff, Jeff
Harvey, Clifford Johnson,
Juan Maldacena,  Shiraz Minwalla, Rob Myers, Joe Polchinski,
and Andy Strominger
for useful discussions.  This work was supported in part by
NSF grant PHY97-22362 to Syracuse University,
the Alfred P. Sloan foundation, and by funds from Syracuse
University.  Much of this work was done while D.M. was a guest of the
Harvard High Energy Group, and he would like to thank them for their
hospitality during this period.



\begin{thebibliography}{99}


\bibitem{SV} A.~Strominger and C.~Vafa,
``Microscopic Origin of the Bekenstein-Hawking Entropy,''
Phys.\ Lett.\  {\bf B379} 99 (1996),
hep-th/9601029.

\bibitem{HW} A. Hanany and E. Witten, {\it ``Type IIB
Superstrings, BPS Monopoles, and Three-Dimensional
Gauge Dynamics,''} Nucl.\ Phys.\  {\bf B492}, 152 (1997), hep-th/9611230.

\bibitem{juan} J. Maldacena, {\it The Large N Limit of
Superconformal Field Theories and Supergravity},
{\bf Adv. Theor. Math. Phys. 2} {\bf 231-252}, hep-th/9711200.

\bibitem{KK} 
R. Sorkin, ``Kaluza-Klein monopole,''
{\it Phys. Rev. Lett.} {\bf 51} 87 (1983);
D. Gross and M. Perry, ``Magnetic Monopoles in Kaluza-Klein
Theories,'' {\it Nucl. Phys.} {\bf B226} 29 (1983).  

\bibitem{ITY} N. Itzhaki, A. Tseytlin, and S. Yankielowicz, {``Supergravity
Solutions for branes localized within branes''} {\it Phys. Lett.}
{\bf B432} 298-304 (1998), hep-th/9803103.

\bibitem{Aki} A. Hashimoto, {``Supergravity solutions for localized
intersections of branes,''} {\it JHEP} {\bf 9901} 019 (1999),
hep-th/9812159.

\bibitem{GM} A. Gomberoff and D. Marolf,
``Brane transmutation in supergravity,''
JHEP {\bf 0002}, 021 (2000),
hep-th/9912184.

\bibitem{Ima} Y. Imamura, {``Supersymmetries and BPS Configurations
on Anti-de Sitter Space,} 
Nucl.\ Phys.\  {\bf B537}, 184 (1999), hep-th/9807179.


\bibitem{CGS} C. G. Callan, A. Guijosa, and K. G. Savvidy, {``Baryons
and String Creation from the Fivebrane Woldvolume Action,''}
Nucl.\ Phys.\  {\bf B547}, 127 (1999),
hep-th/9810092.

\bibitem{NOYY} T. Nakatsu, K. Ohta, T. Yokono, and Y.
Yoshia, {\it ``A proof of Brane Creation via M-theory,''}
hep-th/9711117.

\bibitem{Yosh} Y. Yoshia, {``Geometrical Analysis of Brane
Creation via $M$-theory,''} 
Mod.\ Phys.\ Lett.\  {\bf A13}, 293 (1998), hep-th/9711177.

\bibitem{BG} C. P. Bachas, M. B. Green, {``A Classical Manifestation of the
Pauli Exclusion Principle,''} JHEP {\bf 9801}, 015 (1998), hep-th/9712187.

\bibitem{DBI} A. A. Tseytlin, {``Born-Infeld action, supersymmetry, and
string theory,} hep-th/9908105.  

\bibitem{Thor} L. Thorlacius, ``Born-Infeld String as a Boundary Conformal
Field Theory,'' Phys. Rev. Lett. 80 (1998) 1588-1590,
hep-th/9710181.               

\bibitem{PS} J. Polchinski and A. Strominger, {``New Vacuua for Type
II String Theory,''} 
Phys.\ Lett.\  {\bf B388}, 736 (1996), hep-th/9510227.

\bibitem{Lif} G. Lifschytz, {\it Comparing D-branes to Black Branes}, 
Phys.\ Lett.\  {\bf B388}, 720 (1996),
hep-th/9604156.

\bibitem{BDG} C. P. Bachas, M. R. Douglas, and M. B. Green, 
JHEP {\bf 9707}, 002 (1997),
{``Anomalous Creation of Branes,''} hep-th/9705074.

\bibitem{DFK} U. Danielsson, G. Ferretti, and I. R. Klebanov,
{``Creation of Fundamental
Strings by Crossing D-branes,''} 
Phys.\ Rev.\ Lett.\  {\bf 79}, 1984 (1997),
hep-th/9705084.

\bibitem{BGL} O. Bergman, M. R. Gaberdiel, and G. Lifschytz,
{\it ``Branes, Orientifolds, and the Creation of Elementary
Strings,''} 
Nucl.\ Phys.\  {\bf B509}, 194 (1998), hep-th/9705130.

\bibitem{K} I. R. Klebanov, {``D-branes and the Creation of Strings,''} 
Nucl.\ Phys.\ Proc.\ Suppl.\  {\bf 68}, 140 (1998),
hep-th/97091160.

\bibitem{dA} S. P. de Alwis, {``A note on brane creation,''} 
Phys.\ Lett.\  {\bf B413}, 49 (1997),
hep-th/9706142.

\bibitem{HoWu} P. Ho and Y. Wu, {\it Brane Creation in M(atrix)
Theory,''} 
Phys.\ Lett.\  {\bf B420}, 43 (1998), hep-th/9708137.

\bibitem{WB} E. Witten, {``Anti De-Sitter Space and Holography,''} 
Adv.\ Theor.\ Math.\ Phys.\  {\bf 2}, 253 (1998),
hep-th/9802150.

\bibitem{BGS} C. P. Bachas, M. B. Green, and A. Schwimmer,
{``(8,0) Quantum Mechanics and symmetry enhancement in type I'
superstrings,''}, JHEP {\bf 9801}, 006 (1998), hep-th/9712086.

\bibitem{LeaI}  M. Billo', P. Di Vecchia, M. Frau, A. Lerda, 
I. Pesando, R. Russo, and S. Sciuto, ``Microscopic string analysis of the 
D0-D8 brane system and dual R-R states,''
Nucl.\ Phys.\ {\bf B526} (1998) 199-228, hep-th/9802088.

\bibitem{LeaII} M. Billo', P. Di Vecchia, M. Frau, A. Lerda, and R. Russo,
``The Lorentz force between D0 and D6 branes in String and M(atrix)
theory'', Mod.\ Phys.\ Lett.\ {\bf A13} (1998) 2977, hep-th/9805091.


\bibitem{OSZ} N. Ohta, T. Shimizu, and J-G Zhou,
{\it ``Creation of Fundamental String in M(atrix) Theory,''}
Phys.\ Rev.\  {\bf D57}, 2040 (1998),
hep-th/9710218.

\bibitem{Romans} L. Romans, {\it Phys. Lett.} {\bf B169} 374 (1986).

\bibitem{BDS} C. Bachas, M. Douglas, and C. Schweigert, {``Flux
Stabilization of D-branes,''} JHEP {\bf 0005}, 048 (2000),
 hep-th/0003037.

\bibitem{Taylor} W. Taylor, {\it ``D2-branes in B fields''}, 
hep-th/0004141.

\bibitem{Mor} 
A.~Alekseev, A.~Mironov and A.~Morozov,
``On B-independence of RR charges,''
hep-th/0005244.

\bibitem{3Q} D. Marolf, ``Chern-Simons terms and the Three Notions of Charge,''
hep-th/0006117.

\bibitem{SS} S. Stanciu, ``A Note on D0-branes in Group Manifolds: Flux
Quantization and D0-charge,'' hep-th/0006145.

\bibitem{EGKRS} S. Elitzur, A. Giveon, D. Kutasov, E. Rabinovici, and
G. Sarkissian, ``D-Branes in the Background of NS Fivebranes,''
hep-th/0005052.

\bibitem{Pelc} O. Pelc, ``On the Quantization Constraints 
for a D3 Brane in the Geometry of NS5 Branes,'' hep-th/0007100.

\bibitem{Page} D. N. Page, {it Phys. Rev.} {\bf D 28},
2976 (1983).

\bibitem{CM} C. G. Callan and J. M. Maldacena, {``Brane Death and Dynamics From
the Born-Infeld Action,''}. Nucl.\ Phys.\  {\bf B513}, 198 (1998),
hep-th/9708147.

\bibitem{EWI} E. Witten, 
``Small Instantons in String Theory,'' Nucl.Phys. B460, 541-559 (1996),
hep-th/9511030. 

\bibitem{GP} E. Gimon and J. Polchisnki, 
``Consistency Conditions for Orientifolds and D-Manifolds,''
Phys.\ Rev.\  {\bf D54}, 1667 (1996),
hep-th/9601038.

\bibitem{GKMT} A.~Gomberoff, D.~Kastor, D.~Marolf and J.~Traschen,
Phys.\ Rev.\  {\bf D61}, 024012 (2000),
hep-th/9905094.

\bibitem{Stelle} K. S. Stelle, {\it ``BPS branes in supergravity,''}
hep-th/9803116.

\bibitem{Seiberg}
N.~Seiberg,
``IR dynamics on branes and space-time geometry,''
Phys.\ Lett.\  {\bf B384}, 81 (1996),
hep-th/9606017.

\bibitem{multi} S. Hawking, Phys.\ Lett. {\bf 60A}, 81 (1977).

\bibitem{T-dual} 
J.~Polchinski,
``TASI lectures on D-branes,''
hep-th/9611050; 
J.~Dai, R.~G.~Leigh and J.~Polchinski,
Mod.\ Phys.\ Lett.\  {\bf A4}, 2073 (1989).

\bibitem{Bus} T. Buscher, ``Path Integral Derivation of Quantum Duality
in Nonlinear Sigma Models," {\it Phys. Lett.} {\bf B201} 466 (1988);
``A Symmetry of the String Background Field Equations," {\it Phys. Lett.}
{\bf B194} 59 (1987).   

\end{thebibliography}
\end{document}